# Pair Dispersion in Turbulence

Mickaël Bourgoin[†,1], Nicholas T. Ouellette[†], Haitao Xu[†], Jacob Berg[†,2] & Eberhard Bodenschatz[†,‡]

[†]*Laboratory of Atomic and Solid State Physics, Cornell University, Ithaca, New York 14853, USA;* [‡]*Max Planck Institute for Dynamics and Self-Organization, Göttingen, Germany*

[1]Present Address: Laboratoire des Écoulements Géophysiques et Industriels – C.N.R.S., BP 53-38041 Grenoble Cedex 9, France

[2]Present Address: Risø National Laboratory, DK-4000 Roskilde, Denmark

**Turbulent mixing[1] of liquids and gasses is ubiquitous in nature. It is the basis of all industrial fluid mixing processes, and it determines the spread of pollutants or bioagents in the atmosphere[2] and oceans[3]. Biological organisms even use it to survive in marine ecosystems[4-6]. A fundamental component of turbulent mixing is the separation of two nearby fluid elements, *i.e.*, pair dispersion. Despite almost eighty years of intense scientific inquiry[2,7-16], no clear understanding of this fundamental aspect of turbulence has emerged. One critical unresolved question is the extent to which the initial separation of the fluid particles influences their subsequent motion. Surprisingly, our measurements in a laboratory water flow[17,18] at very high turbulence levels (Taylor microscale Reynolds numbers up to $R_\lambda = 815$) suggest that the initial separation remains important for all but the most violent flows on Earth. This observation has important consequences for such varied problems as pollution control, combustion modelling, hazardous chemical control, and even the understanding of how animals locate food, predators, and mates[5,6].**

For most flows on Earth, both natural and industrial, the turbulence levels are quite small; typically, $R_\lambda < 1000$. Very turbulent atmospheric flows, such as warm clouds or the atmospheric boundary layer[19], have turbulence levels of about $R_\lambda \sim 10^4$. Even the most violent flows of Earth, such as plinian volcanic eruptions, have quite similar turbulence levels. If we approximate a volcano by a turbulent jet with a typical ejection velocity of 100 m/s and an orifice diameter of order 100 m, we find a turbulence level of only about $R_\lambda \sim 10^4$.

In a quiescent fluid, the relative dispersion of two fluid elements (or tracer particles) is dominated by diffusion. The two particles undergo Brownian motion, and the mean square separation between them grows linearly in time. In a turbulent flow, however, if the two particles are separated by distances smaller than the length scale of the largest eddies in the flow, they will separate faster (*i.e.*, superdiffusively). At large separation times and distances, the local correlations responsible for the superdiffusive separation will no longer be present, and, on the average, the relative dispersion will again be linear in time.

We have measured relative dispersion in a water flow at high turbulence levels using optical particle tracking. This technique has been used for a number of years in turbulence research[10,20], but was limited to the measurement of low turbulence level flows due to the fact that tracer particle motions must be resolved over times comparable to the smallest timescale of the flow (i.e., the Kolmogorov time scale $\tau_\eta = (\nu/\varepsilon)^{1/2}$, where $\nu$ is the kinematic viscosity and $\varepsilon$ is the energy dissipation rate per unit mass). In intense turbulence these times are often very small; in our water flow at $R_\lambda = 690$, for example, $\tau_\eta = 0.93$ ms. Previously, using silicon strip detectors from high energy physics[17,18] we extended the particle tracking technique to flows with high turbulence levels. Such detectors, however, are unsuitable for measuring the statistics of many tracer particles at once. Here we use instead Phantom v7.1 digital cameras from Vision Research, Inc. These cameras record 27,000 pictures per second at a resolution of 256×256 pixels. We can use this camera system to track several hundred particles at once[21]. An example of two such simultaneously measured particle tracks is shown in Fig. 1.

By analyzing our measured particle tracks, we have investigated the time evolution of the mean-square separation between two fluid elements. Predictions for the superdiffusivity of this pair dispersion in turbulence date back to 1926 when Richardson[7] suggested that it should grow as $t^3$. By applying Kolmogorov's 1941 scaling theory[22], Obukhov[23] specified that, in the inertial range of turbulence where the only relevant flow parameter is the dissipation rate per unit mass $\varepsilon$, the pair dispersion should grow as $g\varepsilon t^3$, where $g$ is a universal constant. Batchelor[8] in 1950 refined this work, predicting both that the mean-square separation should grow as $t^2$ for short times and that the initial separation should enter the scaling law. Defining $\Delta_i(t)$ as the separation of two fluid elements at time $t$ and along coordinate $i$ and $\Delta_{0i}$ as the initial separation between the fluid elements, Batchelor predicted that, for $\Delta_0$ in the inertial range,

$$\left\langle \left[ \Delta_i(t) - \Delta_{0i} \right]^2 \right\rangle = \frac{11}{3} C_2 \left( \varepsilon \Delta_0 \right)^{2/3} t^2, \quad t < t_0 = \left( \frac{\Delta_0^2}{\varepsilon} \right)^{1/3}, \tag{1}$$

where $C_2$ is the universal constant in the inertial range scaling law for the Eulerian second order velocity structure function with a well-known value of approximately 2.13[24]. Summation is implied over the repeated index $i$. Physically, $t_0$ may be identified as the time for which the two fluid elements "remember" their initial relative velocity, presumably while they are moving in the same eddy of size $\Delta_0$.

To distinguish between the Batchelor scaling and the Richardson-Obukhov scaling, one must have a large inertial range, which implies a large separation between the eddy turnover time $T_L$ and the Kolmogorov time $\tau_\eta$. To achieve such a wide range of scales, the turbulence level must be high, since $R_\lambda \sim (T_L/\tau_\eta)$. Based on evidence from direct numerical simulation (DNS), Yeung[25] has suggested that a turbulence level of at least $R_\lambda = 600$-$700$ is required to see true inertial range scaling of a Lagrangian quantity like the relative dispersion investigated here. Previous experimental and computational studies of dispersion have been limited by their low turbulence levels ($R_\lambda < 300$)[9-14,16]. High turbulence levels are obtained in kinematic simulation models[15], but Thomson & Devenish have recently suggested that such models are ill-suited to the pair dispersion problem[26].

Our measurement of relative dispersion at high turbulence levels is shown in Fig. 2. We find that our data unambiguously scales as $t^2$ for more than two decades in time. This behaviour holds throughout the entire inertial range, even for large initial separations (up to 70% of the energy injection length scale), and for turbulence levels up to $R_\lambda = 815$. We see no Richardson-Obukhov $t^3$ scaling. When we scale our relative dispersion data by the constant predicted by Batchelor, given in eq. 1, the curves collapse onto a single $t^2$ power law. We emphasize that the line drawn in Fig. 2 is not a fit, but rather is exactly $(11/3)C_2(\varepsilon\Delta_0)^{2/3}t^2$. We see some small deviations from this power law for very small initial separations (smaller than roughly 40 times the Kolmogorov length scale $\eta$, the smallest length scale in the flow); this, though, should be expected since these small initial separations do not lie fully in the inertial range.

In Fig. 2, where time is plotted in units of $\tau_\eta$, the data for different initial separations deviate from the $t^2$ law at times that vary with $\Delta_0$. If, however, we scale time by Batchelor's $t_0 = (\Delta_0^2/\varepsilon)^{1/3}$, as shown in Fig. 3, the data for each initial separation deviates from Batchelor's prediction at the same universal value of roughly 0.07 $t/t_0$, irrespective of turbulence levels.

Since our data show that $t_0$ measures the persistence of the Batchelor scaling regime, the existence of a Richardson-Obukhov regime requires not only a large separation between $T_L$ and $\tau_\eta$ but also a large separation between $T_L$ and $t_0$. In our experiments, the maximum value of the ratio of $(T_L/t_0)$ was of order 10, with no hint of $t^3$ scaling. To see a decade of Richardson-Obukhov scaling, then, one must *at minimum* have $(T_L/t_0) \sim 100$. Using the definition of the turbulence level, *i.e.*, the microscale Reynolds number, we can write $R_\lambda = \sqrt{15}\left(T_L/\tau_\eta\right) = \sqrt{15}\left(T_L/t_0\right)\left(\Delta_0/\eta\right)^{2/3}$. To be in the inertial range, the ratio $(\Delta_0/\eta)$ must be at least about $60$[27]. Conservatively, then, we can project from our experiments that a turbulence level of at least $R_\lambda \approx 6000$ is required to see a decade of Richardson-Obukhov scaling. This extrapolation agrees with the prediction from Heppe's stochastic model[28], in which a universal Richardson regime exists for more than a decade in time only at turbulence levels of order $10^4$. Ott & Mann have previously reported a Richardson-Obukhov regime in an experiment with a low turbulence level[10]; as suggested by Sawford, however, this behaviour may be the result of a questionable time shift they applied to their data[29]. Our experiment also allows us to put an upper bound on the initial separations for which a Richardson-Obukhov regime can exist. It is straightforward to show that $(T_L/t_0) = (L/\Delta_0)^{2/3}$ where $L$ is the integral length scale. Then, since we have seen no Richardson-Obukhov scaling for $(T_L/t_0) < 10$, a decade of Richardson-Obukhov scaling requires that $(\Delta_0/L) < 10^{-3}$ irrespective of turbulence level, while still maintaining $(\Delta_0/\eta) > 60$.

An important consequence of these predictions is that in almost all flows with industrial or biological significance the initial separation $\Delta_0$ will influence the subsequent spreading of the two fluid elements throughout the entire period of their turbulent superdiffusive separation. This can explain, for example, the measurements by Warhaft & Lumley[30] of the decay of the fluctuations of a passive scalar injected into the flow. They found that this decay became slower as the separation between two sources was increased. Their results may in turn explain why the spatial arrangement of odour sources plays such an important role in the way crayfish and other crustaceans navigate their marine environments[5].

In summary, we observed that Batchelor's prediction is fulfilled for more than two decades in time at high turbulence levels. While our data may be somewhat contaminated by the inhomogeneity and anisotropy present in our specific flow, the observed scale collapse onto the Batchelor law appears very robust. We saw no Richardson-Obukhov regime up to $R_\lambda = 815$ and suggest that a minimum turbulence level of $R_\lambda \sim 6000$ is required to see a decade of Richardson-Obukhov scaling. Such a high turbulence level is beyond the reach of any current experiments. Our predictions may be tested, however, by the high pressure wind tunnel currently under construction at the Max Planck Institute for Dynamics and Self-Organization. This 1.8 m diameter tunnel will use gaseous sulphur hexafluoride ($SF_6$) pressurised to 15 atmospheres and should be able to reach turbulence levels of approximately $R_\lambda \sim 10^4$ while maintaining measurable length and time scales.

This research is supported by the Physics Division of the National Science Foundation and by the Max Planck Society. We thank L. Collins, J. Hunt, J. Schumacher, D. Vincenzi, and Z. Warhaft for helpful discussions and suggestions over the course of this work.

Correspondence and requests for materials should be addressed to E.B. (e-mail: eb22@cornell.edu).

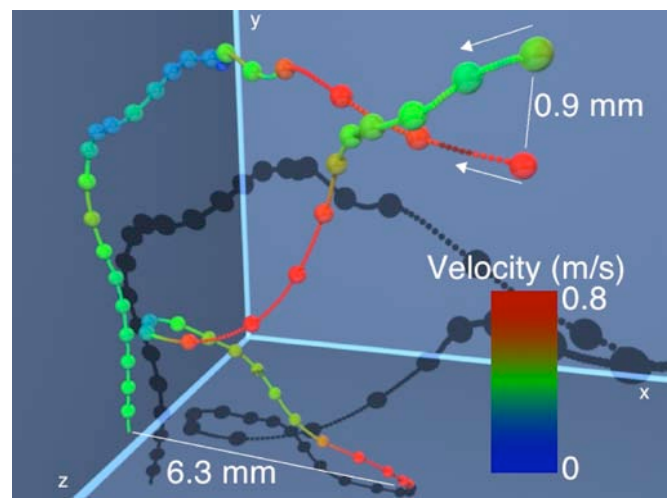

**Figure 1** A pair of measured particle trajectories. Each particle is a polystyrene microsphere with a diameter of 33 $\mu$m and a density of 1.06 g cm$^{-3}$, measured in three dimensions at a Taylor microscale Reynolds number of $R_\lambda$ = 690. The small spheres mark every other measured position of the particles, and are separated by 0.074 ms (≈$\tau_\eta$/13) in time; the large spheres mark every 30th position. The colour of the spheres indicates the magnitude of each particle's absolute velocity in units of m/s. The particles enter the measurement volume as indicated by the arrows, and separate under the influence of the turbulence. We generate turbulence between coaxial counter-rotating baffled disks in a closed chamber with a volume of approximately 0.1 m³. We make measurements in a subvolume of roughly (5 cm)³ in the centre of the tank where the mean flow is statistically zero. We use three Phantom v7.1 cameras arranged in a plane with an angular separation of roughly 45° to track the particles. Each camera records images at a rate of 27,000 frames per second at a resolution of 256×256 pixels, corresponding to 25 measurements per $\tau_\eta$ at $R_\lambda$ = 690 and 15 measurements per $\tau_\eta$ at $R_\lambda$ = 815. The particles are illuminated by two frequency-doubled pulsed Nd:YAG lasers at a wavelength of 532 nm and with a combined power of roughly 130 W. The particle positions are measured with a precision of roughly 0.1 pixels[21], corresponding to about 20 μm in the flow. Energy dissipation rates are determined from measurements of the Eulerian structure functions, and the energy injection scale $L$ is found to be approximately 7 cm. At the highest turbulence level reported in this work, the Kolmogorov length scale $\eta = (\nu^3/\varepsilon)^{1/4}$ is 23 μm and the corresponding time scale $\tau_\eta$ is 0.54 ms. See refs. 17 and 18 for detailed information about this flow.

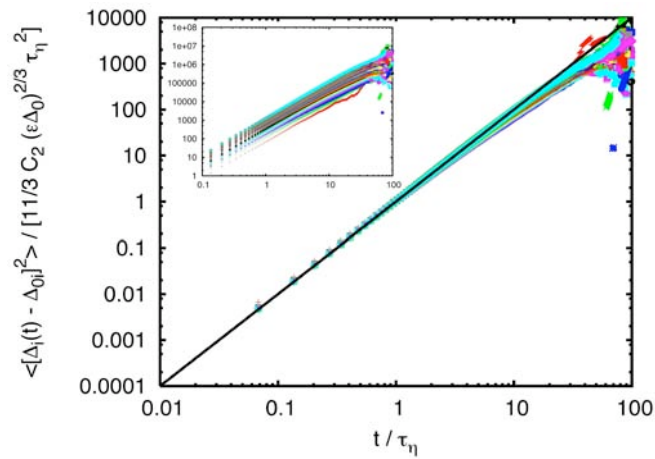

**Figure 2** Evolution of the mean-square particle separation. The mean square separation between particle pairs is plotted against time for fifty different initial separations at a turbulence level of $R_\lambda$ = 815, with both axes normalized by the Kolmogorov scales. Each curve represents a bin of initial separations 1 mm wide (≈43$\eta$), ranging from 0-1 mm to 49-50 mm. The curves are scaled by the constant (11/3)$C_2(\varepsilon\Delta_0)^{2/3}$, defined in eq. 1. As expected, the data collapse onto a single universal power law. The bold line drawn is not a fit to these data, but is instead the exact power law predicted by Batchelor[8]. We note that since the smallest $\Delta_0$ measured is not in the inertial range, we do not expect it to scale perfectly as $t^2$, and indeed it does not scale as well as the larger $\Delta_0$. The inset shows the same curves scaled simply by the Kolmogorov length, for which we see no scale collapse.

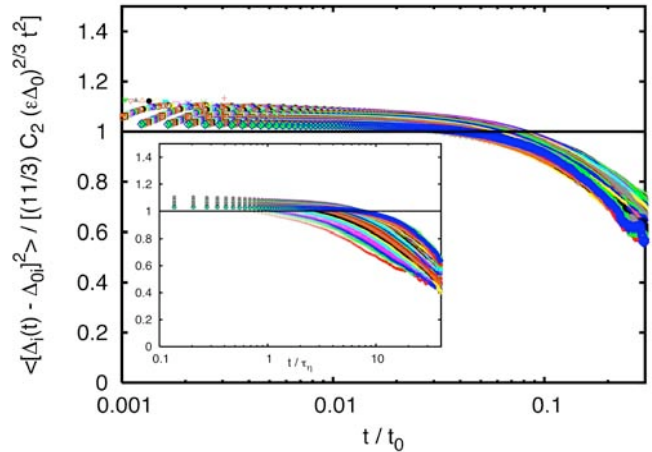

**Figure 3** Mean square separation with time scaled by $t_0$. The mean square separation at $R_\lambda$ = 815 compensated by Batchelor's scaling law (eq. 1) is plotted against time in units of $t_0 = (\Delta_0^2/\varepsilon)^{1/3}$. The inset shows the same compensated data plotted against time scaled by the Kolmogorov time. The data clearly collapses significantly better with time scaled by $t_0$. The data begins to deviate from a $t^2$ power law at a universal time of about 0.07 $t/t_0$.